\documentclass[runningheads]{llncs}
\usepackage[T1]{fontenc}

\usepackage{graphicx}
\usepackage{amsmath,amssymb}
\usepackage{booktabs}
\usepackage[inline]{enumitem}
\usepackage{comment}
\usepackage{multirow}
\usepackage{xspace}
\usepackage{makecell}
\usepackage[table, dvipsnames]{xcolor}

\usepackage[symbol]{footmisc}

\usepackage[pagebackref,breaklinks,colorlinks]{hyperref}
\usepackage[capitalize]{cleveref}
\crefname{section}{Sec.}{Secs.}
\Crefname{section}{Section}{Sections}
\Crefname{table}{Table}{Tables}
\crefname{table}{Tab.}{Tabs.}

\newcommand{\myfirstpara}[1]{\par \noindent \textbf{#1.}}
\newcommand{\mypara}[1]{\vspace{0.1em} \myfirstpara{#1}}

\def\sota{\texttt{SOTA}\xspace}

\def\myarch{\texttt{GLCM-MAE}\xspace}
\def\mae{\texttt{MAE}\xspace}
\def\maes{\texttt{MAEs}\xspace}
\def\glcm{\texttt{GLCM}\xspace}
\def\rgb{\texttt{RGB}\xspace}

\def\mse{\texttt{MSE}\xspace}
\def\ssim{\texttt{SSIM}\xspace}
\def\kde{\texttt{KDE}\xspace}
\def\bx{\mathbf{x}\xspace}
\def\cL{\mathcal{L}\xspace}
\def\cG{\mathcal{G}\xspace}
\def\GU{\textbf{\texttt{GU}}\xspace}
\def\BU{\textbf{\texttt{BU}}\xspace}
\def\PX{\textbf{\texttt{PX}}\xspace}
\def\CC{\textbf{\texttt{CC}}\xspace}

\begin{document}

\title{Focus on Texture: Rethinking Pre-training in Masked Autoencoders for Medical Image Classification}
\titlerunning{Rethinking Pre-training in MAEs}

\author{Chetan Madan\inst{1}\thanks{Joint first authors~} \and
Aarjav Satia\inst{1}\textsuperscript{*} \and
Soumen Basu\inst{1}\thanks{Soumen contributed while he was at IIT Delhi} \orcidID{0000-0002-3915-7545}  \and
Pankaj Gupta\inst{2} \and
Usha Dutta\inst{2} \and
Chetan Arora\inst{1}
}

\authorrunning{C. Madan et al.}

\institute{Indian Institute of Technology, Delhi \and
PGIMER Chandigarh
}

\maketitle              

\begin{abstract}
Masked Autoencoders (\maes) have emerged as a dominant strategy for self-supervised representation learning in natural images, where models are pre-trained to reconstruct masked patches with a pixel-wise mean squared error (\mse) between original and reconstructed \rgb values as the loss. We observe that \mse encourages blurred image reconstruction, but still works for natural images as it preserves dominant edges. However, in medical imaging, when the texture cues are more important for classification of a visual abnormality, the strategy fails. Taking inspiration from \texttt{Gray Level Co-occurrence Matrix} (\glcm) feature in Radiomics studies, we propose a novel \mae based pre-training framework, \myarch, using reconstruction loss based on matching \glcm. \glcm captures intensity and spatial relationships in an image, hence proposed loss helps preserve morphological features. Further, we propose a novel formulation to convert matching \glcm matrices into a differentiable loss function. We demonstrate that unsupervised pre-training on medical images with the proposed \glcm loss improves representations for downstream tasks. \myarch outperforms the current state-of-the-art across four tasks -- gallbladder cancer detection from ultrasound images by 2.1\%, breast cancer detection from ultrasound by 3.1\%, pneumonia detection from x-rays by 0.5\%, and COVID detection from CT by 0.6\%.
Source code and pre-trained models are available at: \\\url{https://github.com/ChetanMadan/GLCM-MAE}.

\keywords{Medical Image classification \and Masked Autoencoders \and Image reconstruction}

\end{abstract}

\section{Introduction}
Masked Autoencoders (\maes) have emerged as a powerful approach for self-supervised representation learning in natural images \cite{he2022masked,wei2023diffusion}. In \mae based pre-training, one masks parts of input image, and then trains a neural network model to reconstruct the masked regions. Importantly, the whole training process does not require any supervised labels, which allows one to use large amount of training images available on the internet, thus helping a model to learn robust image representation from highly varied data. Several researchers have shown that \mae based pre-training followed by fine-tuning on downstream task helps achieve state-of-the-art performance on variety of tasks \cite{adamae,ristea2024self}.  

Whereas, several variations \cite{adamae,wei2025towards,yao2023one} of \mae pre-training have been proposed, all these use pixel-wise mean squared error (\mse) between the original and reconstructed patch to compute the loss, and then back-propagate the loss to train the model. We observe that such pre-training is not effective for medical images. Our analysis reveals that the problem lies in the pixel-wise \mse based reconstruction loss. The particular loss encourages model to generate smooth images which works for contour or dominant edge based classification in natural images. 

In medicine, \texttt{radiomics} is a technique which extracts an array of texture features from medical images. These features, known as radiomic features, are then used to train simple machine learning algorithms like SVMs and decision trees. \glcm is one such radiomic feature which analyzes how often gray levels (intensities) occur together within an image, considering only the neighboring pixels. By counting co-occurrences in a specific pixel offset (often 1 or 2), \glcm effectively captures the spatial arrangement of textures. In this work we propose a novel pre-training framework for medical images, which uses \mae with \glcm guided reconstruction loss. This enables a model to focus on subtle gray-level variations within each patch, and better preserve morphological features.

\mypara{Contributions} 
%
\begin{enumerate*}[label=\textbf{(\arabic*)}]
    \item We investigate the reasons behind relative ineffectiveness of the \mae based pre-training on medical imaging tasks. We observe that pixel-wise \mse based reconstruction over-smooth subtle, but semantically important, texture features in the image representation, and degrade performance on medical image classification. 
    \item We propose a novel \mae technique tailored for medical imaging, incorporating a \glcm-guided loss. \glcm is non-differentiable. However, we do a clever work-around based on a differentiable joint histogram of an image, shifting the image by one pixel to generate a \glcm-like representation. This design ensures that the learned representations preserve critical morphological information, which is essential for downstream tasks such as classification and segmentation in medical imaging.
    \item We demonstrate the efficacy of \myarch with the proposed \glcm loss, outperforming existing \sota on four medical image analysis tasks: gallbladder cancer detection from ultrasound by 2.1\%, breast cancer detection from ultrasound by 3.1\%, pneumonia detection from x-rays by 0.5\%, and COVID detection from CT by 0.6\%.
    \item We are releasing source-code and pre-trained models for the community.
\end{enumerate*}

\section{Proposed Method}
\myfirstpara{\mse Loss}
Let $\bx$ denote the original image, and $\bx_{m}$ the masked patches. \mse loss, $\cL$, for \rgb-based reconstruction is computed as:
$\cL = \frac{1}{N} \sum_{i=1}^{N} \| \bx_{m}^{(i)} - \hat{\bx}_{m}^{(i)} \|_2^2$.    
Here, $N$ is the number of masked patches, $\bx_{m}^{(i)}$ is the ground truth \rgb value of the $i$-th masked patch, and $\hat{\bx}_{m}^{(i)}$ is the reconstructed \rgb value. 

\mypara{Revisiting Gray-Level Co-Occurrence Matrix (\glcm)}
Let $i$ and $j$ represent pixel intensity values having $G$ gray levels (e.g. for 8-bit grayscale image, $G=256$), $d$ is the distance between the pixel pairs, and $\theta$ is the angle (e.g., $0^\circ$, $45^\circ$, $90^\circ$, or $135^\circ$). \glcm is a $G \times G$ matrix, denoted by $\cG(v, w \mid d, \theta)$, where each element $(v,w)$ represents the number of times a pixel with intensity $v$ is paired with a pixel of intensity $w$ at the given distance and direction. Given a grayscale patch $i$ with pixel intensities $I_i(x, y)$ where $x, y$ represent pixel locations, the \glcm matrix records the co-occurrence frequency of intensity values across pixel pairs within a defined spatial offset $d$. We denote the co-occurrence count for a pair of intensity values $(v, w)$ as:
\begin{equation}
\cG^{(i)}(v,w) = \sum_{(x,y), (x',y')} \delta(I_i(x, y) = v) \cdot \delta(I_i(x', y') = w),
\end{equation}
where $\delta$ is the Kronecker delta function that is 1 if its argument is true, and 0 otherwise. Additionally, $(x', y')$ represent the neighbor pixel of $(x,y)$.

\begin{figure}[t]
	\centering
	\includegraphics[width=0.8\linewidth]{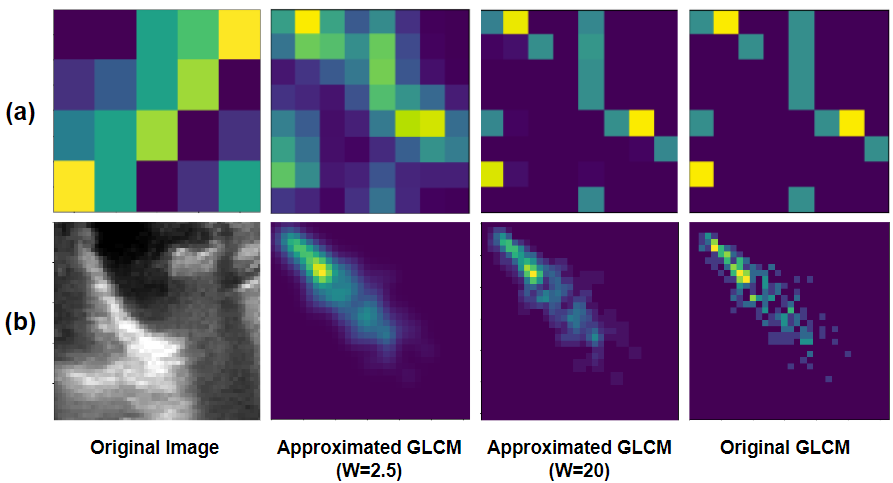}
	\caption{Differentiable approximate GLCM construction converges to original GLCM with sufficiently large kernel bandwidth ($W$). Row (a) contains a sample image, followed by the \glcm approximations for different values of $W$, and the original \glcm. Row (b) shows an image patch from the GBCU dataset along with the \glcm approximations and original \glcm.}
	\label{fig:glcm_approx_vis}
\end{figure}

\mypara{Differentiable Approximation of \glcm}
Note that, \glcm is non-differentiable. Hence, we compute a differentiable approximation of \glcm matrix for each masked patch independently. We use Kernel Density Estimation (\kde) technique proposed by \cite{avi2023differentiable} to compute differentiable histograms. Let $I_i \in [0, 255]$ denote a gray level value of an image pixel $i$. \kde for estimating the intensity density $f_I$ of an image $I$ with $N$ pixels is given by: $\hat{f}_I(x) = \frac{1}{NW} \sum_{i=1}^N \mathcal{K} \left( \frac{I_i - x}{W} \right)$, where $x \in [0,255]$, $\mathcal{K}(\cdot)$ denotes the kernel, and $W$ is the bandwidth. We select the kernel as the derivative of sigmoid function: $\mathcal{K}(z) = \sigma'(z) = \sigma(z)\sigma(-z)$ where $\sigma(z) = \frac{1}{1 + e^{-z}}$ is the sigmoid function. 
For a given patch $I$, we calculate two offset images -- $I_1 =$ dropping the last column of $I$, and $I_2 = $ dropping the first column of $I$. Following \cite{avi2023differentiable}, we calculate a differentiable joint histogram between $I_1$ and $I_2$ as: $H(I_1, I_2) = \frac{1}{N} P_1 P_2^T$, where, the \kde-based probabilities $P_1$ and $P_2$ for each intensity bin $k$ are defined as:
\begin{equation}
\resizebox{0.93\hsize}{!}{
	$ 
	P_1 = \sigma\left(\frac{I_1 - \mu_k + L/2}{W}\right) - \sigma\left(\frac{I_1 - \mu_k - L/2}{W}\right), \quad
	P_2 = \sigma\left(\frac{I_2 - \mu_k + L/2}{W}\right) - \sigma\left(\frac{I_2 - \mu_k - L/2}{W}\right)
	$.
}
\end{equation}
Here, $\mu_k$ is the center of the $k^{\text{th}}$ intensity bin, $W$ is the bandwidth which controls the smoothness, $L = \frac{2}{K}$ is the length of bins, where $K$ is the total number of bins. The sigmoid function, creates a soft, continuous approximation of bin boundaries.
For sufficiently large bandwidth $W$, $H(I_1, I_2)$ converges to a smooth approximation of the \glcm. This smooth joint histogram, illustrated in \cref{fig:glcm_approx_vis}, retains key structural properties of the \glcm while being differentiable.
To extend this approach vertically, we repeat the above formulation to compute $V(I_1, I_2)$, representing the \glcm calculated from vertically adjacent pixel pairs. This allows for directional, and differential \glcm computation compatible with gradient-based learning.

\begin{figure*}[t]
	\centering
	\includegraphics[width=0.9\linewidth]{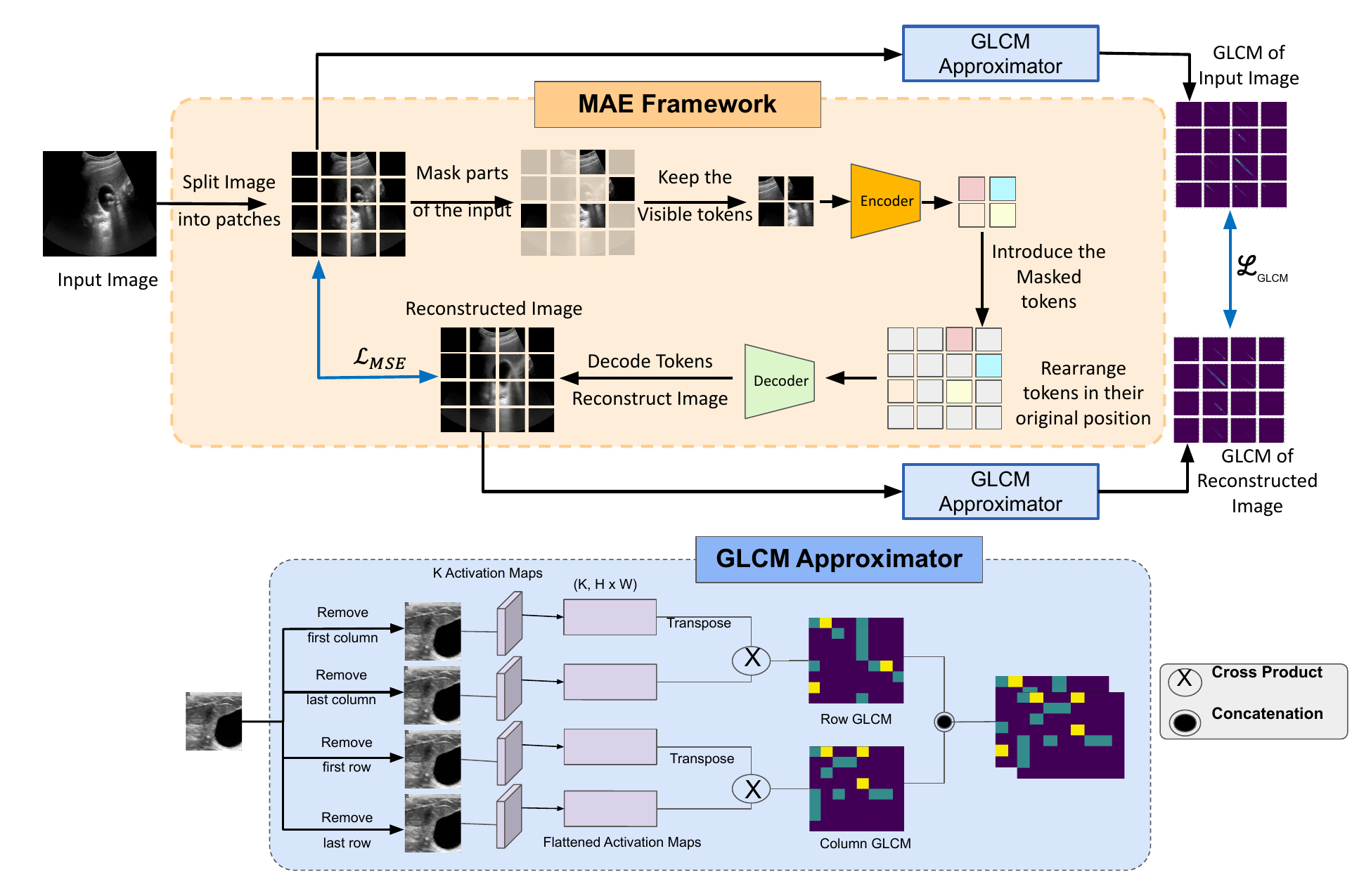}
	\caption{Overview of the proposed \myarch pipeline. We propose novel \glcm loss guided reconstruction for improved representation learning in medical imaging tasks. The \textit{black lines} represent forward pass and the \textit{\textcolor{blue}{blue lines}} represent loss computation.}
	\label{fig:arch}
\end{figure*}

\mypara{\glcm Loss Function}
Once the normalized \glcm is obtained, we calculate the \glcm loss by comparing the predicted and ground-truth \glcm matrices for each patch. Let $\hat{\cG}_p^{(i)}$ and $\hat{\cG}_g^{(i)}$ denote the differentiable \glcm computed for predicted and ground truth patch $i$. The \glcm loss is computed as the mean squared error between these matrices:
\begin{equation}
\cL_{\text{GLCM}} = \frac{1}{|M|} \sum_{i \in M}\sum_{v, w} \left( \hat{\cG}_p^{(i)}(v,w) - \hat{\cG}_g^{(i)}(v,w) \right)^2
\label{eq:loss}
\end{equation}
where $M$ refers to the set of masked patches. 

\mypara{Adding \mse and \ssim Loss for Stable Pre-training}
While the \glcm loss helps capture subtle intensity relationships, it lacks global consistency due to its focus on local intensity relationships. Hence we further add pixel-wise \mse loss, and Structural Similarity Index (\ssim) loss to preserve visual structures in an image. We define the final loss function as:
$\cL = \alpha \cL_{\text{MSE}} + \beta \cL_{\text{GLCM}} + \gamma \cL_{\text{SSIM}}$,
where $ \alpha $, $ \beta $, and $ \gamma $ are scaling factors. During the later phase of training, we scale down $\alpha $ to 0.1 to reduce the dominance of \mse loss, and preserve morphological features due to \glcm and \ssim losses.
\cref{fig:arch} shows the proposed pipeline.

\subsection{\textbf{\myarch} Architecture}
\cref{fig:arch} shows a brief depiction of the proposed \myarch framework. Below we discuss the different parts of the proposed architecture.

\mypara{Masked Token Generation}
Similar to vanilla \mae, we divide the image into patches of $16\times16$ pixels and randomly mask 75\% of the patches using uniform sampling. We choose a high masking ratio of 75\% to prevent any data leakage while reconstructing the masked patches.

\mypara{Encoder}
We use a standard ViT architecture with embedding dimension of 768, encoder depth of 12 layers and 12 heads. Similar to the vanilla \mae strategy, only the visible image patches are passed to the encoder and the masked patches are dropped at the encoder to allow the encoder to train effectively on a fraction of the full image.

\mypara{Decoder}
The encoded visible patches, along with the masked patches are fed to the decoder, which is a shallow ViT and has a depth of 8, number of heads as 16 and embedding dimensions as 512. The decoder is only used during the \myarch pre-training and not during the finetuning of the downstream task.

\mypara{Warm-up Phase}
Training is divided into two phases. In the initial 200 epochs, we use only \mse reconstruction loss to provide stable gradients and improve structural learning ($\beta\!=\!0, \gamma\!=\!0$). This phase helps the model establish a coherent base representation, reducing instability in later training.

\mypara{MAE Pre-training Phase}
In the next 200 epochs, we activate the \glcm and \ssim losses, with the \mse loss scaled down by a factor of 0.1 ($\alpha\!=\!0.1, \beta\!=\!1, \gamma\!=\!1$). This phase promotes learning of both subtle intensity patterns (via \glcm) and structural integrity (via \ssim).

\begin{table}[t]
	\caption{(Left) The 5-fold cross-validation (Mean) accuracy, specificity, and sensitivity of baselines and \myarch in detecting GBC from the US. \myarch achieves the best accuracy and sensitivity, which is much desired for GBC detection. (Right) The AUROC, Accuracy, and F1 of baselines and \myarch for identifying Pneumonia from chest X-Ray images. We report the metrics on the test set consisting of 390 pneumonia and 234 normal images.
	}
	\label{tab:gbcu}
	\centering
	\resizebox{\linewidth}{!}{%
		\begin{tabular}{|lccc|p{1cm}|lccc|}
			\cline{1-4} \cline{6-9}
			\textbf{Method} & \textbf{Acc.} &  \textbf{Spec.} & \textbf{Sens.} 
			& & 
			\textbf{Method} & \textbf{AUROC} &  \textbf{Acc.} & \textbf{F1.}
			\\
			\cline{1-4} \cline{6-9} 
			%
			ResNet50 \cite{resnet} & 0.867 & 0.926 & 0.672 & & VAE \cite{kingma2013auto} & 0.618 & 0.640 & 0.774\\

        VGG16 \cite{vgg} & 0.693 & 0.960 & 0.495 & & GANomaly \cite{akcay2018ganomaly} & 0.780 & 0.700 & 0.790\\
        
        InceptionV3 \cite{inception} & 0.844 & 0.953 & 0.807 & & f-MemAE \cite{gong2019memorizing} & 0.778 & 0.565 & 0.826 \\
        
        RetinaNet \cite{retinanet} & 0.749 & 0.867 & 0.791 & & MNAD \cite{Park_2020_CVPR} & 0.773 & 0.736 & 0.793 \\
        
        EfficientDet \cite{efficientdet} & 0.739 & 0.881 & 0.858 & & IGD \cite{chen2022deep} & 0.734 & 0.740 & 0.809 \\
        
        ViT \cite{vit} & 0.803 & 0.901 & 0.860 & & SQUID \cite{Xiang_2023_CVPR} & 0.876 & 0.803 & 0.847\\
        
        DEIT \cite{touvron2021training} &  0.829  & 0.900 & 0.875 & & CheXzero \cite{tiu2022expert} & 0.927 & 0.830 & 0.875 \\
        
        PVTv2 \cite{wang2021pvtv2} & 0.824 & 0.887 & 0.894 & & Xplainer \cite{pellegrini2023xplainer} & 0.899 & 0.782 & 0.850 \\     
    	GBCNet \cite{basu2022surpassing} & 0.921 & \textbf{0.967} & 0.919 & & CoOp \cite{zhou2022learning} & 0.946 & 0.846 & 0.886 \\
        US-USCL \cite{basu2022unsupervised} & 0.921 & 0.926 & 0.900 & & MaPLe \cite{khattak2023maple} & 0.951 & 0.894 & 0.895 \\
        RadFormer \cite{basu2023radformer} & 0.921 & 0.961 & 0.923 & & PPAD \cite{Sun_PositionGuided_MICCAI2024} & \textbf{0.967} & 0.894 & 0.918 \\
        \cline{1-4} \cline{6-9}
        MAE \cite{he2022masked} & 0.928 & 0.937 & 0.921 & & MAE \cite{he2022masked} & 0.941 & 0.949 & 0.922\\
        OmniMAE \cite{omnimae} & 0.855 & 0.927 & 0.621 & & OmniMAE \cite{omnimae} & 0.953 & 0.950 & \textbf{0.968}  \\
        \cline{1-4} \cline{6-9}
        \myarch (Ours) & \textbf{0.949}& 0.957 & \textbf{0.934} & & \myarch (Ours) & 0.964 & \textbf{0.955} & 0.964 \\
        \cline{1-4} \cline{6-9}
    \end{tabular}
	}
\end{table}

\begin{table}[!t]
	\caption{The 5-fold cross validation (Mean$\pm$SD) accuracy, specificity, and sensitivity of baselines and \myarch in detecting breast cancer from ultrasound data.}
	\label{tab:busi}
	\centering
    \small
    \resizebox{\linewidth}{!}{%
    \begin{tabular}{lccccccc}
        \Xhline{3\arrayrulewidth}
        & ResNet50\cite{resnet} & DenseNet121\cite{densenet} & RadFormer\cite{basu2023radformer} & MAE\cite{he2022masked} & OmniMAE\cite{omnimae} & \textbf{Ours} \\
        \Xhline{\arrayrulewidth}
        Accuracy & - & - & 0.842$\pm$0.063 & 0.892$\pm$0.013 & 0.832$\pm$0.050 & \textbf{0.923$\pm$0.014} \\
        Specificity & \textbf{0.975$\pm$0.024} & 0.968$\pm$0.018 & 0.881$\pm$0.069 & 0.945$\pm$0.019 & 0.947$\pm$0.034 & 0.936$\pm$0.029 \\
        Sensitivity & 0.829$\pm$0.088 & 0.824$\pm$0.027 & 0.738$\pm$0.105 & 0.748$\pm$0.059 & 0.614$\pm$0.124 & \textbf{0.852$\pm$0.028} \\
        \Xhline{3\arrayrulewidth}
    \end{tabular}}
\end{table}

\begin{figure*}[t]
	\centering
	\includegraphics[width=0.92\textwidth]{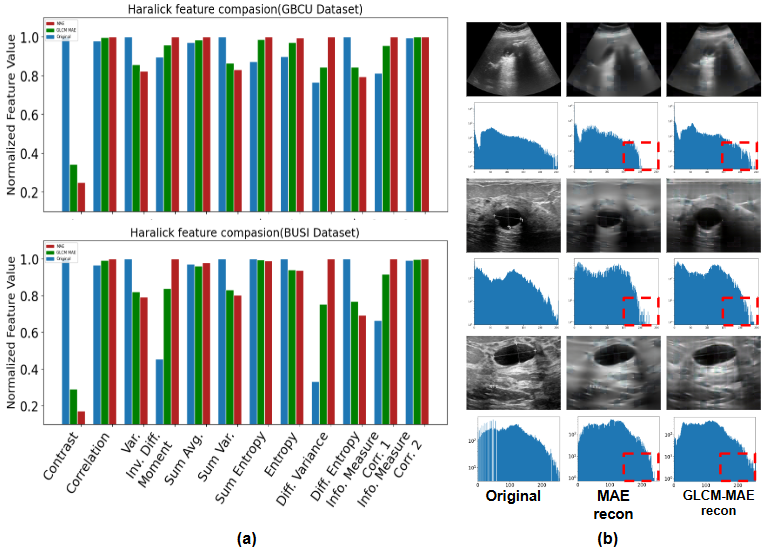}
	\caption{\textbf{(a)} Shows the Haralick feature values derived from Original image (in blue), \myarch reconstructions (in green) and \mae reconstructions (in red) averaged over the test entire set of GBCU and BUSI dataset. The plot suggests better texture preservation of \myarch reconstruction relative to \mae over 12 texture based Haralick features. \textbf{(b)} Shows histogram comparisons between reconstructions of \mae and \myarch for images from GBCU and BUSI dataset, clearly depicting the loss of subtle gray level intensity values in \mae reconstructions, which are better conserved by \myarch. }
	\label{fig:glcm_vis}
\end{figure*}

\section{Experiments and Results}
\mypara{Datasets Used}
We use four problems/datasets to demonstrate the efficacy of our approach, viz,
\begin{enumerate*}[label=\textbf{(\arabic*)}]
\item [\GU:] gallbladder cancer detection from ultrasound  \cite{basu2022surpassing,basu2023gall}, 
\item [\BU:] breast cancer detection from ultrasound  \cite{al2020dataset}, 
\item [\PX:] pneumonia detection from chest X ray \cite{chestxray}, and 
\item [\CC:] COVID-19 detection from lung CT-scan dataset \cite{Soares2020.04.24.20078584}.
\end{enumerate*}
Due to restriction on paper length, we skip the problem/dataset details and request the reader to refer to the respective original papers.

\mypara{Implementation and Evaluation}
\begin{enumerate*}[label=\textbf{(\arabic*)}]
\item \textbf{Pretraining with \myarch:} 
We implemented our experiments using PyTorch. The pretraining is done in two stages: we first train an MAE with ViT-B backbone using pixel-wise \mse reconstruction loss for 200 epochs. For the second stage, we resume training the model with a combination of the proposed loss in \cref{eq:loss} for 200 epochs. For both stages, we mask 75\% patches and train using a learning rate of $10^{-6}$ with weight decay of $0.05$. We use a batch size of 8 and AdamW optimizer for pretraining.
\item \textbf{Finetuning for Downstream Tasks:} 
We finetune our pre-trained ViT encoder for 100 epochs for downstream classification task. The learning rate starts from $10^{-3}$ and decays to a minimum value of $10^{-6}$ with a weight decay of $0.05$ with AdamW optimizer and a batch size of 128. We select the model that gives the highest accuracy on the validation set.
\item \textbf{Evaluation Metrics:} 
We used accuracy, specificity (true negative rate), and sensitivity (true positive rate/ recall) for GU, and BU, AUROC for CC, and AUROC and F1-score for PX evaluation. 
\end{enumerate*}

\mypara{Efficacy of \textbf{\myarch} over \sota Baselines}
To validate the effectiveness of our \myarch, we apply it on four disease detection tasks: \GU, \BU, \PX, and \CC. Our experiments reveal that \myarch beats the task specific \sota on all four datasets. We have used \sota mask-based representation learning baselines namely, OmniMAE (CVPR2023) \cite{omnimae} along with the original MAE \cite{he2022masked} for comparison with our proposed \myarch. Note that the MAE baseline methods use \mse loss based reconstruction in contrast to our \glcm-guided loss. 
\textbf{Quantitative Results: }
We show the quantitative evaluation results in \cref{tab:gbcu,tab:busi}. Additionally, we also report the results on COVID classification using CT images, where we achieve an AUROC of 99.19, exceeding the current SOTA, CropMixPaste\cite{Self-Supervised(TMI-2024)} by 7.8\% and exceed other MAE variants’ \cite{he2022masked} performance by 0.6\%.
\textbf{Statistical Significance: }
The p-values of \myarch vs. MAE (\GU= 0.036; \BU= 0.005; \PX= 0.034; \CC= 0.048) show that \myarch gains are significant.

\begin{figure*}[t]
	\centering    
    \includegraphics[width=0.8\linewidth]{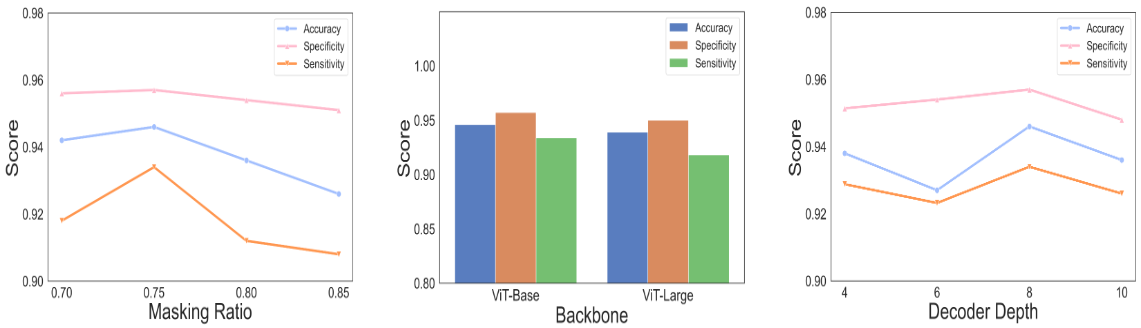}
	\caption{Ablation study. We report the mean scores over 5-fold cross-validation for GBC detection. (a) Effect of varying the masking ratio. (b) Using different encoder backbones. (c) Effect of varying the decoder depth.}
	\label{fig:ablation}
\end{figure*}

\mypara{Masked Patch Reconstruction Analysis}
We extract the Haralick features from the original, and reconstructed images using \mae and \glcm visually compare the outputs (\cref{fig:glcm_vis}). We see that reconstructions from \glcm have Haralick features similar to the original, providing conclusive evidence of \glcm's superior performance. 
The \myarch reconstructions are visually closer to the original image, particularly in regions with subtle texture and intensity variations. 

\begin{table}[t]
	    \caption{Ablation Study Results. (Left) Effect of the kernel bandwidth $W$ value on GBCU data. (Right) Effect of different reconstruction loss functions on \myarch on the GBCU Dataset.}
	\label{tab:ablation_combined}
	
	\centering
	\resizebox{\linewidth}{!}{%
		\begin{tabular}{lcccp{1cm}lccc}
			\cline{1-4} \cline{6-9}
		        \textbf{Bandwidth ($W$)} & \textbf{Acc.} & \textbf{Spec.} & \textbf{Sens.} & & \textbf{Loss Function} & \textbf{Acc.} & \textbf{Spec.} & \textbf{Sens.} \\
			\cline{1-4} \cline{6-9} 
			$W=5$  & 0.939 & 0.941 & 0.924  & & MSE  & 0.928  & 0.937 & 0.919 \\

                $W=15$  & 0.940  & 0.947  & 0.924  & & GLCM & 0.930 & 0.947 & 0.919   \\
        
        $W=30$  & \textbf{0.949} & \textbf{0.957} & \textbf{0.934} & & GLCM + MSE & 0.939 & 0.950 & 0.919  \\
        \cline{1-4} 
        & & & &&  GLCM + SSIM  & 0.914 & 0.944 & 0.852  \\
        & & & && GLCM + MSE + SSIM & \textbf{0.949} & \textbf{0.957} & \textbf{0.934} \\
			\cline{6-9}
		\end{tabular}
	}
\end{table}

\mypara{Ablation Study on the GBCU dataset}
\begin{enumerate*}[label=\textbf{(\arabic*)}]
\item \textbf{Selection of Reconstruction Loss Function: }
\cref{tab:ablation_combined}(right) shows the effect of choosing the loss functions: Mean Squared Error (\mse), \glcm loss, Structural Similarity Index (\ssim), and their combinations, on the quality of the representation learned for the downstream GBC detection task.
\item \textbf{Bandwidth Selection: }
\cref{tab:ablation_combined}(left) shows the effect of different kernel bandwidths ($W$) for GLCM computation. Kernel size of $W\!=\!30$  offers a minor improvement over lower kernel sizes. 
\item \textbf{Masking Ratio: }
\cref{fig:ablation} (a) compares different masking ratios for \myarch. We found that a masking ratio of 0.75 achieved the best balance between learning generalizable features and preserving sufficient image context.
\item \textbf{Encoder: }
\cref{fig:ablation} (b) shows the effect of choosing different Vision Transformer (ViT) variants as encoders for the token encoding task.
\item \textbf{Decoder Depth: }
We conduct experiments by adjusting the decoder blocks, with results shown in \cref{fig:ablation}(c). Performance improves when the decoder depth increases from 4 to 8. However, performance declines when the depth is extended to 10, aligning with findings in \cite{basu2024focusmae,madan2025lq}. 
\end{enumerate*}

\section{Conclusion}
In this work, we introduced \myarch, a novel texture-aware self-supervised pre-training framework tailored for medical image classification. Unlike traditional \maes relying on pixel-wise \mse-based reconstruction, we integrate a \glcm-guided loss to capture essential textures and spatial relationships in medical images. With \glcm guidance, our model effectively learns subtle intensity variations that are often missed in \mse based reconstructions, enhancing feature extraction suited for medical imaging needs. In summary, \myarch sets a benchmark for texture-sensitive self-supervised learning, opening avenues for texture-based losses in medical imaging applications.

\begin{credits}
\subsubsection{\ackname} 
\footnotesize We thank the funding support from the Department of Biotechnology (grant number BT/PR51120/AI/133/179/2024), and Ministry of Education funding (grant number IITJMU/CPMU-AI/2024/0002) for the Centre of Excellence for AI in healthcare. Chetan Arora's travel is supported by Yardi School of AI Travel Grant.

\subsubsection{\discintname}
The authors have no competing interests to declare that are relevant to the content of this article. 

\end{credits}
{\small
\bibliographystyle{splncs04}
\bibliography{main}
}

\end{document}